\documentclass[twocolumn, 10pt]{article}
\usepackage{amsmath,amssymb}

\usepackage{calligra}
\DeclareMathAlphabet{\mathcalligra}{T1}{calligra}{m}{n}
\DeclareFontShape{T1}{calligra}{m}{n}{<->s*[2.2]callig15}{}
\newcommand{\scriptr}{\mathcalligra{r}\,}

\newcommand{\scriptt}{\mathcalligra{t}\,}

\usepackage{graphicx}
\usepackage{float}
\usepackage{diagbox}
\usepackage[format=hang,labelfont=bf,textfont=small,singlelinecheck=false,justification=raggedright,margin={12pt,12pt},figurename=Fig.]{caption}
\usepackage{titlesec}
\usepackage{mathtools}
\usepackage{textcomp}
\usepackage{cite}
\usepackage{float}
\usepackage{hyperref}
\usepackage{subcaption} 
\usepackage[export]{adjustbox}
\usepackage{etoolbox}

\newcommand{\lambdabar}{{\mkern0.75mu\mathchar '26\mkern -9.75mu\lambda}}

\makeatletter
\def\affiliation#1{\gdef\@affiliation{#1}}
\def\abstract#1{\gdef\@abstract{#1}}
\def\graphabst#1{\gdef\@graphabst{#1}}
\def\keywords#1{\gdef\@keywords{#1}}
\def\corresp#1{\gdef\@corresp{#1}}

\newcommand{\MakeTitle}{
  \newpage
  \null
  \vskip 2em%
  \begin{center}%
  \Large \@title\par
  \vskip 1em%
  \large \@author
  \end{center}
  \noindent\@affiliation\par
  \vskip 1em%
  \noindent\@corresp\par
  \vskip 1em%
  \noindent\@abstract\par
  \vskip 1em%
  \noindent\@keywords\par
}
\makeatother

\makeatletter
\patchcmd{\@maketitle}{\raggedright}{\centering}{}{}
\patchcmd{\@maketitle}{\raggedright}{\centering}{}{}
\makeatother

\newcommand*{\TitleFont}{%
      \usefont{\encodingdefault}{\rmdefault}{}{n}%
      \fontsize{18}{12}%
      \selectfont}

\titleformat{\section}
  {\normalfont\fontsize{10}{11}\bfseries}{\thesection.}{2pt}{}
  \titlespacing*{\section}{0pt}{12pt}{6pt}

\titleformat{\subsection}
  {\normalfont\fontsize{10}{10}\bfseries}{\thesubsection.}{2pt}{}
  \titlespacing*{\subsection}{0pt}{6pt}{0pt}
  
\titleformat{\subsubsection}
  {\normalfont\fontsize{10}{10}\bfseries}{\thesubsubsection.}{2pt}{}
  \titlespacing*{\subsubsection}{0pt}{6pt}{0pt}
  
\normalsize
\title{\TitleFont 
A complement to the scalar wave theory of light}
\author{Kolahal Bhattacharya
}

\affiliation{
\begin{center}
Homi Bhabha Centre for Science Education (TIFR), Mumbai-400088
\end{center}
}

\corresp{
}

\abstract{\textbf{Abstract}: 
In this paper, we discuss how the concepts of Hamiltonian optics are internally connected to the scalar wave theory of light rays. It is shown that the solutions of the reduced wave equation are similar to Huygen's wavelets, and they can be used to understand Snell's law of refraction. This model can also be used to derive the coefficient of reflection consistent with Fresnel's equation for $s$-polarized light.}


\begin{document}
\onecolumn
\MakeTitle
\section{Introduction}

Interdisciplinary approaches are common in science and technology, and their importance in pedagogy cannot be overstated. Williams et al.~\cite{williams2003aren } analyse the reasons why many high school students do not find physics interesting and suggest that an interdisciplinary approach might enhance students' interests in physics. A new idea can be better appreciated by the students when they interpret it in terms of the known concepts bearing no apparent connections with the new topic. Apart from understanding concepts in a new light, it often leads to important milestones. For example, the discovery of the double helix structure of DNA required the knowledge of chemistry, molecular biology, and crystallography. This trend has been adopted in the pedagogy of physics curricula at the undergraduate level in recent decades. An example of this is `F=ma optics'~\cite{evans1986f} where many elements of geometrical optics were interpreted in terms of the concepts learned from Newtonian mechanics. This analogy was a glimpse of the deeper connection between classical mechanics and ray optics through the principle of stationary action~\cite{ goldstein2002classical}. Whereas in classical mechanics, this principle helps in understanding the dynamical evolution of a system, in geometrical optics it is used to analyse the spatial evolution of the light ray in a medium with a known refractive index. This approach culminated in a new discipline called  Hamiltonian optics~\cite{buchdahl1993introduction, torre2005linear, dragoman2013quantum, lakshminarayanan2002lagrangian} and is not usually taught in undergraduate-level physics courses. The present author thinks that concrete examples of such interdisciplinary approaches could be very useful for the students.

The analytical formulation of geometrical optics raises the question of whether it is possible to rediscover the quantum nature of light in an appropriate limit starting from this alternative approach to optics. In fact, the answer is affirmative: Gloge and Marcuse developed a quantum theory of light rays~\cite{gloge1969formal} based on Hamiltonian optics. This work finds the emergence of the quantum effects in the limit where the wavelength of light rays cannot be neglected and eventually derives the reduced wave equation (which they called the Klein Gordon equation of the quantum theory of light rays) whose solutions could represent the probability wave amplitudes of the light rays.

In the introductory level courses on optics, students learn about Huygen's principle~\cite{jenkins2018fundamentals}, to understand the propagation of light rays in a medium and Snell's law of light refraction between media with different refractive indices without any reference to the quantum description of light. With higher courses on electromagnetic theory, they also learn about the coefficient of reflection and transmission when light, an electromagnetic wave, falls at the interface of two media~\cite{hecht1998hecht}. A few unconventional approaches to refraction were illustrated in the pedagogical works, by (a) Drosdoff et al.~\cite{drosdoff2005snell} who treated photons as entities with well-defined energy and momenta, to prove Snell's law, (b) Evans et al.~\cite{evans1986f} who argued that the tangential component of the optical equivalent of mechanical velocity must be the same in two media. Another method has been demonstrated by Ghatak~\cite{lakshminarayanan2002lagrangian} who set the derivative of total optical path length to zero to arrive at Snell's law. The last example made use of Hamilton's principle of stationary action. It is not clear if these methods could develop to the point where one can obtain the coefficients of reflection and transmission. These different approaches do not use the quantum description of light rays, but we believe that the use of this description may bring pedagogic insight to these topics.

In particular, we did not find any instance where the reduced wave equation of the light rays introduced in the formal quantum theory of light rays~\cite{gloge1969formal} was utilised to prove Snell's law or to obtain the coefficients of the light reflection or refraction. However, such a development would be pedagogically very appealing, because the solutions of this equation are very similar to the Huygen's wavelets. In this article, we will present an algebraic method to fill in this gap. The students do not need prior knowledge of Hamiltonian optics, but some familiarity with  analytical mechanics and Schr$\rm{\ddot o}$dinger's equation is helpful. It is hoped that the present manuscript could be a valuable addition to the existing literature of physics pedagogy.

We will start our discussion with a discussion on the known concepts and published materials in next section~\ref{Sec2} which provides a summary of Hamilton's formulation of geometrical optics and leads to the differential equation describing the spatial evolution of the light rays. Evans et al.~\cite{evans1986f} presented a clever parametrization that significantly reduces the degree of complexity of this ray equation. In the next section~\ref{ScWvEq}, we shall find that the reduced wave equation which was developed by Gloge et al.~\cite{gloge1969formal} nicely reflects the elements of the formulation~\cite{evans1986f}. Later in that section, the main results of this paper will be presented. It will be observed that this equation can be interpreted as a Schr$\rm{\ddot{o}}$dinger's equation with zero energy wavefunctions. Using a model of incidence of the light rays (represented by such wavefunctions) on a potential barrier in two dimensions, we will obtain Snell's law of refraction. We will see that this approach also leads to Fresnel's equations~\cite{hecht1998hecht} for the $s$-polarized light. We shall conclude with some discussion on the implication of this study.

\section{Analytical formulation of optics}\label{Sec2}
\subsection{Lagrangian formulation}
Hamilton's principle of stationary action is introduced to the students in analytical mechanics courses. Action $S$ of a dynamical system is defined as the time integral of Lagrangian density $\mathcal{L}=\mathcal{L}(q,\dot{q},t)$:\\
\begin{equation}\label{Eq1}
    S=\int_{t_1}^{t_2}\mathcal{L}(q,\dot{q},t)\ dt
\end{equation}
Sticking to the usual convention, we shall call $\mathcal L$ the Lagrangian for its subsequent appearances. For the fixed time instants $t_1$ and $t_2$, the principle says that the system will dynamically evolve along a path, for which the action will be stationary to the first order. That is, the action will not change due to a slight variation of the path. This statement is mathematically expressed as:
\begin{equation}\label{Eq2}
\delta S=\delta \int_{t_1}^{t_2}\mathcal{L}(q,\dot{q},t)\ dt=0
\end{equation}
The above principle leads to the Euler-Lagrange equation that governs the dynamical evolution of the system:
\begin{equation}\label{Eq3}
    \frac{\partial\mathcal{L}}{\partial q} =\frac{d}{dt}\left(\frac{\partial\mathcal{L}}{\partial\dot{q}}\right)
\end{equation}
\subsection{Fermat's principle}
In case of geometrical optics, the equivalent physical principle is the Fermat's principle which states that for fixed end points a ray of light will propagate along a trajectory along which the optical path length $\mathbb{L}=\int n\ ds$ is stationary. That means the variation in optical path length will not change to the first order:
\begin{equation}\label{Eq4}
    \delta\mathbb{L}=\delta\int_{P_1}^{P_2} n\ ds=0
\end{equation}
Euler-Lagrange variation of the principle (Eq.\eqref{Eq4}) leads to the differential ray equation governing the spatial evolution of the light rays:
\begin{equation}\label{Eq5}
    \frac{\partial{n}}{\partial{\bf r}} =\frac{d}{ds}\left(n\frac{d{\bf r}}{ds}\right)
\end{equation}
This equation is non-linear and is difficult to solve when the refractive index $n$ varies spatially.
\subsection{F=ma optics}
Evans~\cite{evans1986f} showed that if Fermat's principle is expressed in terms of an independent variable $a$, defined by $n=\left|\frac {d{\bf r}}{da}\right|$, where $n$ denotes the refractive index and ${\bf r}$ denotes the position vector, then Eq.\eqref{Eq5} can be expressed as:
\begin{equation}\label{Eq6}
\frac{d^2{\bf r}}{da^2}=-\nabla^2\left(\frac{n^2}{2}\right)
\end{equation}
-where $a$ is a parameter. This equation has the well-known form of Newton's second law of motion, in terms of the parameter $a$. In this formulation, the quantity that is equivalent to force is proportional to the gradient of the square of the refractive index. This specialized coordinate is applicable to geometrical optics, in which mass, velocity, kinetic and potential energies do not have their standard units. This correspondence is shown below in table~\ref{T1}.
\begin{table}[ht]
 \begin{center}
 \renewcommand{\arraystretch}{1.5}
 \begin{tabular}{||c | c | c||}
  \hline
         			& Definitions in classical mechanics	& Equivalent quantities in geometrical optics\\\hline
Position		&	   ${\bf r}(t)$		&	${\bf r}(a)$	\\\hline
time			&			t			&		a			\\\hline
velocity		&	$\frac{d{\bf r}}{dt}\equiv{\bf{\dot r}}$			&		$\frac{d{\bf r}}{da}\equiv{\bf r}'$			\\\hline
potential energy&		$U({\bf r})$		&		$-\frac{n^2({\bf r})}{2}$			\\\hline
mass			&		$m$		&			$1$		\\\hline
kinetic energy	&		$T=\frac{m}{2}\left|\frac{d{\bf r}}{dt}\right|^2$		&		$\frac{1}{2}\left|\frac{d{\bf r}}{da}\right|^2$		\\\hline
total energy	&		$\frac{m}{2}\left|\frac{d{\bf r}}{dt}\right|^2+U({\bf r})$		&			$\frac{1}{2}\left|\frac{d{\bf r}}{da}\right|^2-\frac{n^2}{2}=0$\\\hline
 \end{tabular}
 \end{center}
\caption{Mechanical view of optics}\label{T1}
\end{table}
From table~\ref{T1}, we find that the optical equivalent of potential energy is a quadratic function of refractive index and the equivalent of total energy (Hamiltonian) is 0. 

\subsection{Hamiltonian optics}
A similar formulation of the problem is possible in terms of the optical Hamiltonian, expressed as a function of $(x, y, z)$, and their conjugate momenta $p_x$, $p_y$ and $p_z$. Here we take $b$ as a stepping parameter along the ray. Then, the expression for optical Lagrangian becomes:
\begin{equation}\label{Eq7}
    L(x,y,z,x',y',z',b)=n(x,y,z)\sqrt{x'^2+y'^2+z'^2}
\end{equation}
-where $x'=\frac{dx}{db}$ etc. The conjugate momenta are defined as:
\begin{align}\label{Eq8}
    p_x=\frac{\partial L}{\partial x'}=n\frac{x'}{\sqrt{x'^2+y'^2+z'^2}}=n\frac{dx}{ds}=n_x\nonumber\\
    p_y=\frac{\partial L}{\partial y'}=n\frac{y'}{\sqrt{x'^2+y'^2+z'^2}}=n\frac{dy}{ds}=n_y\nonumber\\
    p_z=\frac{\partial L}{\partial z'}=n\frac{z'}{\sqrt{x'^2+y'^2+z'^2}}=n\frac{dz}{ds}=n_z,
\end{align}
where $(n_x,n_y,n_z)$ denote the components of refractive index vector ${\bf n}=(n_x,n_y,n_z)$. The Hamiltonian, constructed by Legendre transformation, can be evaluated to be equal to zero\footnotemark[1]:
\begin{equation}\label{Eq9}
    H(x,y,z,p_x,p_y,p_z,b)=x' p_x+y' p_y +z' p_z - L=0
\end{equation}
\footnotetext[1]{We comment that if $z$ coordinate were taken as the independent variable, instead of $b$, then we would have the momenta $p_x$ and $p_y$, conjugate to the coordinates $x(z)$ and $y(z)$. In that case, Hamiltonian would be expressed as: $H=-\sqrt{n^2-p_x^2-p_y^2}$}. 
\subsubsection{Quantum description of light rays}
From the ray view of light, one can make the transition to the quantum description of light by treating momenta by appropriate linear differential operators: $p_x\rightarrow\hat{p}_x\equiv{-i\frac {\lambda}{2\pi}\frac{\partial}{\partial x}}$ etc. These act upon wavefunctions $\psi$ that represent light rays via the eigenvalue equations:
\begin{equation}\label{Eq9a}
    \hat{p}_x\psi=n_x\psi{\hspace{1.0cm}}\implies{\hspace{1.0cm}\hat{p}^2_x\psi=n_x^2\psi}
\end{equation}
Repeating this exercise for all three components, and summing up the squared equations, we find the reduced wave equation: 
\begin{equation}\label{Eq9b}
    \lambdabar^2\nabla^2\psi+n^2\psi=0
\end{equation}
Quantum effect manifests in the limit $\lambdabar\not\rightarrow 0$, and geometrical optics emerges in the limit when $\lambdabar \rightarrow 0$. Gloge et al.~\cite{gloge1969formal} used $z$ as the stepping parameter to arrive at this equation. On the other hand, we chose $b$ as the stepping parameter to show the internal consistency with the formulation presented in `F=ma' optics. 

We note that the so-called `quantum' description of light rays is equivalent to the scalar wave description of light, usually taught at the undergraduate level as wave optics.

\subsection{Scalar wave description}
The electromagnetic theory asserts that light is a travelling transverse electromagnetic wave in which the electric field ${ \vec E}({\bf r},t)$ and the magnetic field ${\vec B}({\bf r},t)$ are coupled in the form of a pulse (see section 3.2 of~\cite{hecht1998hecht}). The spatio-temporal variation of the electric field in a region devoid of source electric charge and current can be expressed by the vector wave equation:
\begin{equation}\label{Eq9c}
\nabla^2{\vec E}-\frac{1}{v^2}\frac{\partial^2\vec{E}}{\partial t^2}=0,
\end{equation}
where $v$ denotes the speed of light in the medium of  propagation. Each component of the electric field vector $E_i\equiv(E_x, E_y, E_z)$ denotes a pulse that satisfies the scalar wave equation:
\begin{equation}\label{Eq9d}
    \nabla^2{E_i}-\frac{1}{v^2}\frac{\partial^2{E_i}}{\partial t^2}=0
\end{equation}
If we assume that the electric pulse $E_i({\bf r},t)$ can be expressed as a product of spatial part $u(\bf r)$ and temporal part $T(t)$, then using the separation of the variables technique, we can deduce the differential equation for the spatial part of the light wave:
\begin{equation}\label{Eq9e}
    \nabla^2 u({\bf r})+k^2u({\bf r})=0
\end{equation}
-where $k$ is a constant representing the wavenumber. This equation is called the reduced wave equation or the Helmholtz equation. One can also express this equation in the form:
\begin{equation}\label{Eq10}
    \lambdabar^2\nabla^2 u({\bf r})+n^2u({\bf r})=0,
\end{equation}
where $n$ represents the refractive index of the medium and $\lambdabar=\frac{\lambda}{2\pi}$ is the reduced wavelength of light waves in this medium. Clearly, Eq.\eqref{Eq9b} and Eq.\eqref{Eq10} are identical, and thus $\psi$ -representing a ray of light, and $u$ -representing a pulse of the electric field, must be intimately related. 

\section{Reduced wave equation as Schr$\rm{\ddot{o}}$dinger's equation}\label{ScWvEq}
We note that Eq.\eqref{Eq9b} can also be expressed as:
\begin{equation}\label{Eq11}
    -\frac{\lambdabar^2}{2\times 1}\nabla^2 \psi - \frac{n^2}{2} \psi = 0\cdot \psi
\end{equation}
Comparing Eq.\eqref{Eq11} with the generic form of the time independent Schr$\rm{\ddot{o}}$dinger's equation and with the terms in table~\ref{T1}, we observe that the optical equivalent of the Hamiltonian for light rays is $0$ and the optical equivalent of the potential energy is $-\frac{n^2}{2}$. Of course, the dimension of the Hamiltonian (in Eq.\eqref{Eq9}) is not of energy. Gloge et al.~\cite{gloge1969formal} (1969) could not identify Eq.\eqref{Eq10} as a Schr$\rm{\ddot{o}}$dinger's equation for the light rays\footnotemark[2], 
\footnotetext[2]{In fact, they derived an approximate form of Schr$\rm{\ddot{o}}$dinger's equation in the paraxial approximation where the wavefunctions are non-zero energy states.}
since they could not identify the `mass term' for light rays as 1, `total energy' as $0$ and the `potential energy term' as $-\frac{n^2}{2}$, that were accomplished by the authors of `F=ma optics' in 1986.

\subsection{Connection with Huygen's wavelets and phasors}
The elementary eigensolutions\footnotemark[3] of Eq.\eqref{Eq9b} in a homogeneous medium are given by~\cite{torre2005linear} (a) plane waves $\left(e^{i{\bf k}\cdot{\bf r}}\right)$, and (b) spherical waves $\left(\frac{\lambda}{r}e^{i{\bf k}\cdot{\bf r}}\right)$, for $r>\lambda$. In a typical wave optics experiment taught in the undergraduate physics curriculum, \footnotetext[3]{One can construct the general travelling wave solutions of the wave equation~\eqref{Eq9d} by superposing the incoming and outgoing waves. In one and three dimensions, the general solution can be expressed as $E_x(x,t)=f(x+vt)+g(x-vt)$, and $E_r(r,t)=\frac{F}{r}(r+vt)+\frac{G}{r}(r-vt)$, respectively.} one shines an aperture (or a slit) of dimension $a$ with monochromatic light of wavelength $\lambda$, and a screen is placed on the opposite side, at a distance $L$ from the  aperture, to observe possible diffraction patterns. Assuming that the monochromatic light incident on the aperture can be assumed to be plane waves, the Fresnel number of the optical setup can be defined as~\cite{hecht1998hecht}:
\begin{equation}
    \mathcal{F}=\frac{1}{L}\left(\frac{a^2}{\lambda}\right)
\end{equation}
If the observation screen is close enough to the aperture, in a way such that $\mathcal F\gtrsim1$, then one is working in the near field limit. In this case, it is convenient to use the spherical wave basis to describe the near field diffraction. However, if it is far away, such that $\mathcal{F}<<1$, then one is working in the far field limit. In this limit, one can use the plane wave solution, as locally spherical wavefronts behave as plane waves. 
\begin{figure}[ht]
\centering
\hspace{-1.0 cm}
\begin{subfigure}{.50\textwidth}
  \centering
  \captionsetup{justification=centering}
  \includegraphics[height=5.0 cm, width= 5.5 cm]{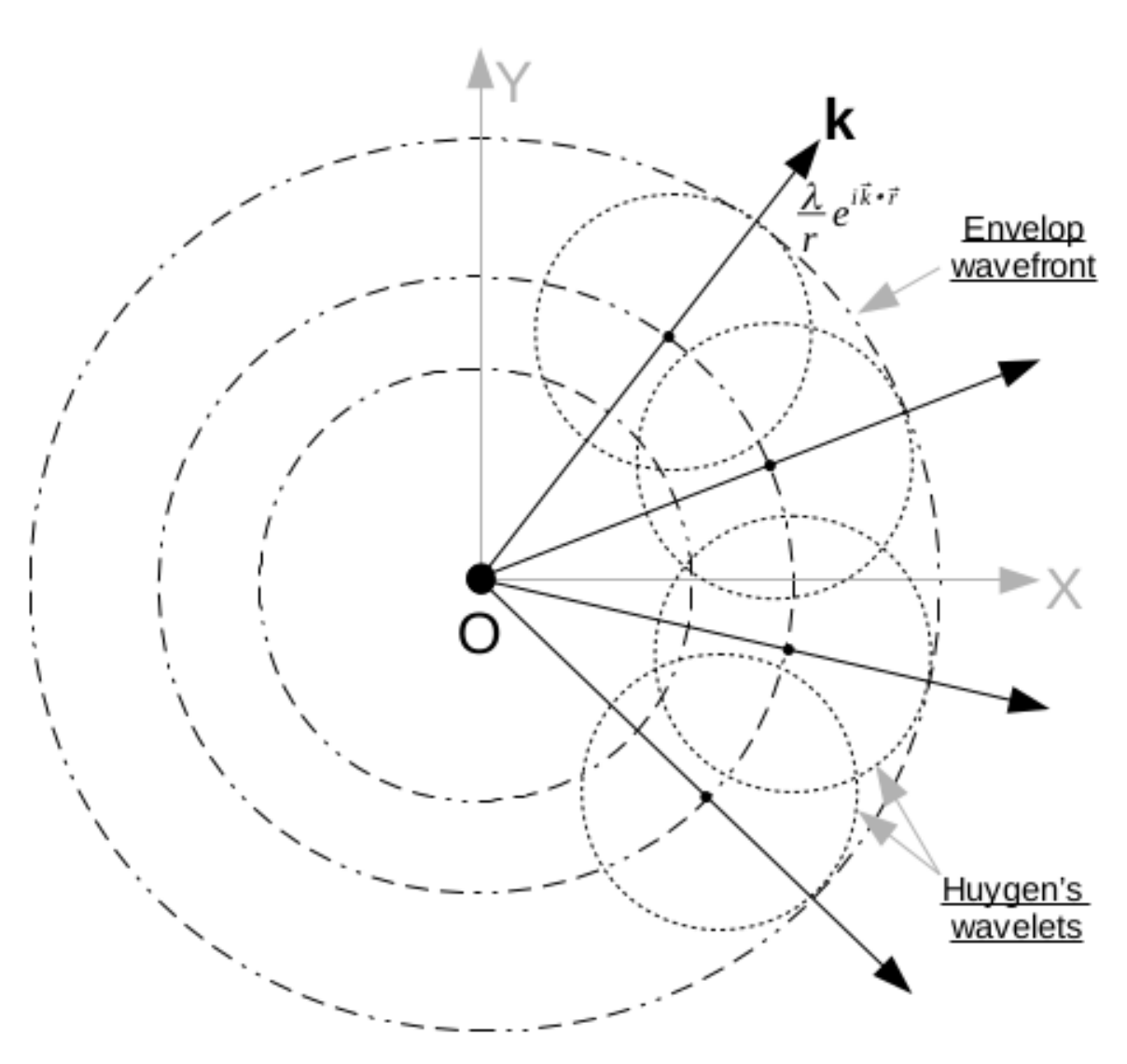}
  \caption{}
  \label{fig0a}
\end{subfigure}
\hspace{-1.0 cm}
\begin{subfigure}{.15\textwidth}
  \centering
  \captionsetup{justification=centering}
  \includegraphics[height=5.5 cm, width= 3.0 cm]{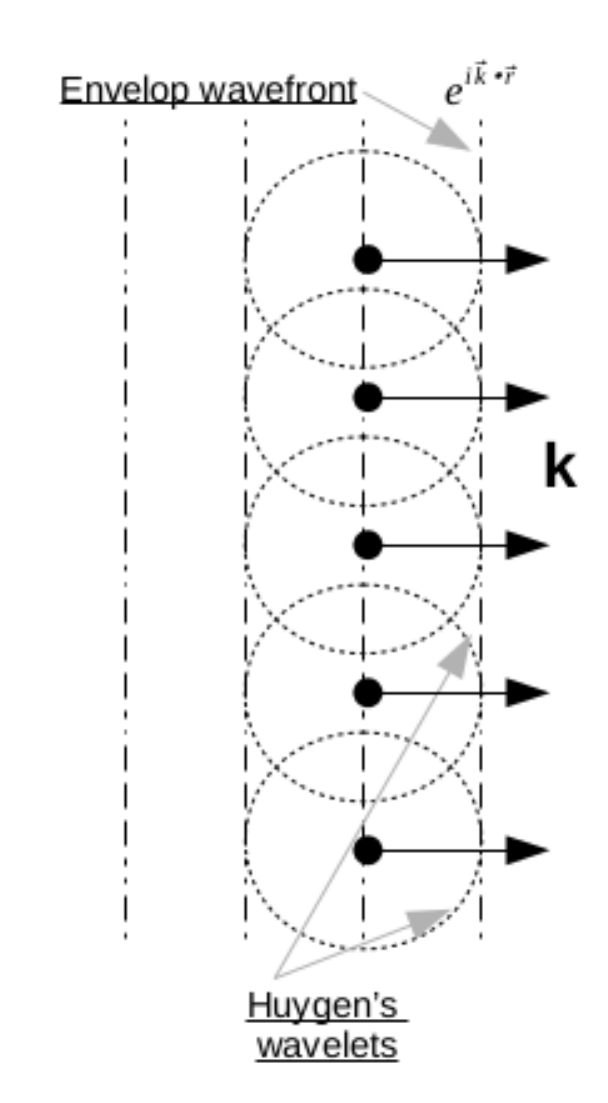}
  \caption{}
  \label{fig0b}
\end{subfigure}
\hspace{0.5 cm}
\begin{subfigure}{.35\textwidth}
  \centering
  \captionsetup{justification=centering}
  \includegraphics[height=3.0 cm, width= 5.0 cm]{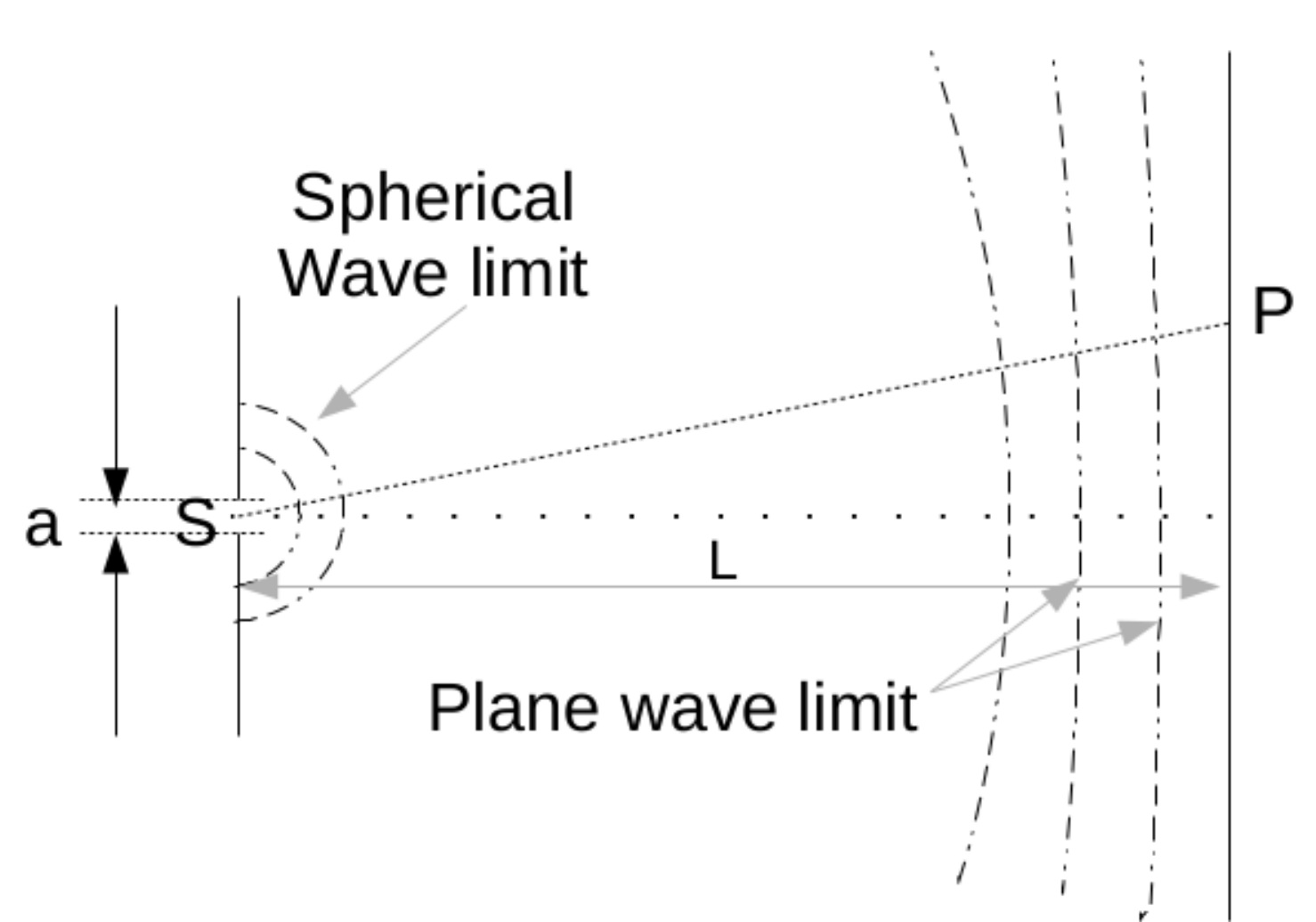}
  \caption{}
  \label{fig0c}
\end{subfigure}
\caption{Propagation of (a) a spherical wave, and (b) a plane wave. The wavefronts at $t+\Delta t$ are envelops of spherical wavelets generated from the wavefronts at $t$. (c) In far the field limit, the spherical wavefronts locally behave as the plane waves. Mathematical functions of these envelops are solutions of Eq.\eqref{Eq9b}.} 
\label{fig0}
\end{figure}

These wavefronts may indeed be considered as the envelopes of Huygen's wavelets, generated from the wavefront at a previous step of light propagation. In the far field limit, these wavelets are just the plane waves and can be expressed as: $\psi =Ae^{i{\bf k}\cdot{\bf r}}$ where $A$ is a complex constant (the initial phase for the monochromatic wave is assumed to have been absorbed within $A$). As such, there is no problem with the complex wavefunction, but this cannot represent a real-valued pulse of the electric field. The electric field pulse can be written as $u=Re\left[Ae^{i{\bf k} \cdot{\bf r}}\right]$. These pulses are referred to as phasors, due to the phase contained in the exponent (note that ${\bf k} \cdot{\bf r}$ is a phase). This observation can also be taken as a justification for taking the electric field pulse $u$ of Eq.\eqref{Eq10} as a complex quantity, commonly practised to simplify calculations. The phasor addition of the waves is utilised in the derivation of the intensity patterns of many  interference and diffraction experiments~\cite{halliday2010physics}.

\subsection{Particle (or light ray) incident on potential barrier} \label{Sec3}
The discussion in the previous section~\ref{ScWvEq} throws light on the scalar wave nature of the optical field. Specifically, we found that the zero energy wavefunctions of light rays satisfy Schr$\rm{\ddot o}$dinger's equation. Therefore, it might be possible to derive Snell's law of geometrical optics. To check this, we begin with the problem of a particle or a ray of light of energy $E$, in the region with potential $V_0$ at $x<0$, incident obliquely on a potential barrier $V_1$ at $x\ge0$\footnotemark[4]. We represent this particle (or ray of light) by the wavefunction $\psi$. In the context of geometrical optics, such a potential barrier arises due to the difference of the refractive indices of the two media, as shown in the following figure~\ref{fig1}. 
\footnotetext[4]{The energy and momentum of a ray of light are understood in the sense of table~\ref{T1}.}

\begin{figure}[ht]
\centering
\begin{subfigure}{.45\textwidth}
  \centering
  \captionsetup{justification=centering}
  \includegraphics[height=5.0 cm, width= 7.5 cm]{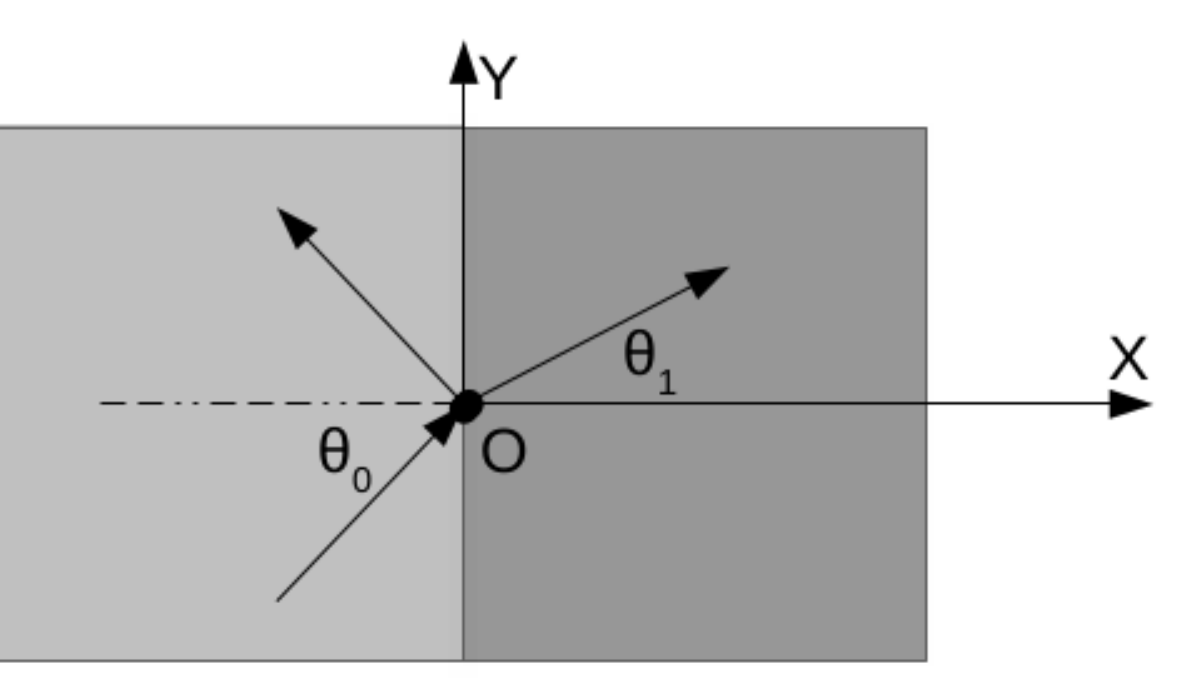}
  \caption{}
  \label{fig1a}
\end{subfigure}
\hspace{0.0 cm}
\begin{subfigure}{.45\textwidth}
  \centering
  \captionsetup{justification=centering}
  \includegraphics[height=5.0 cm, width= 7.5 cm]{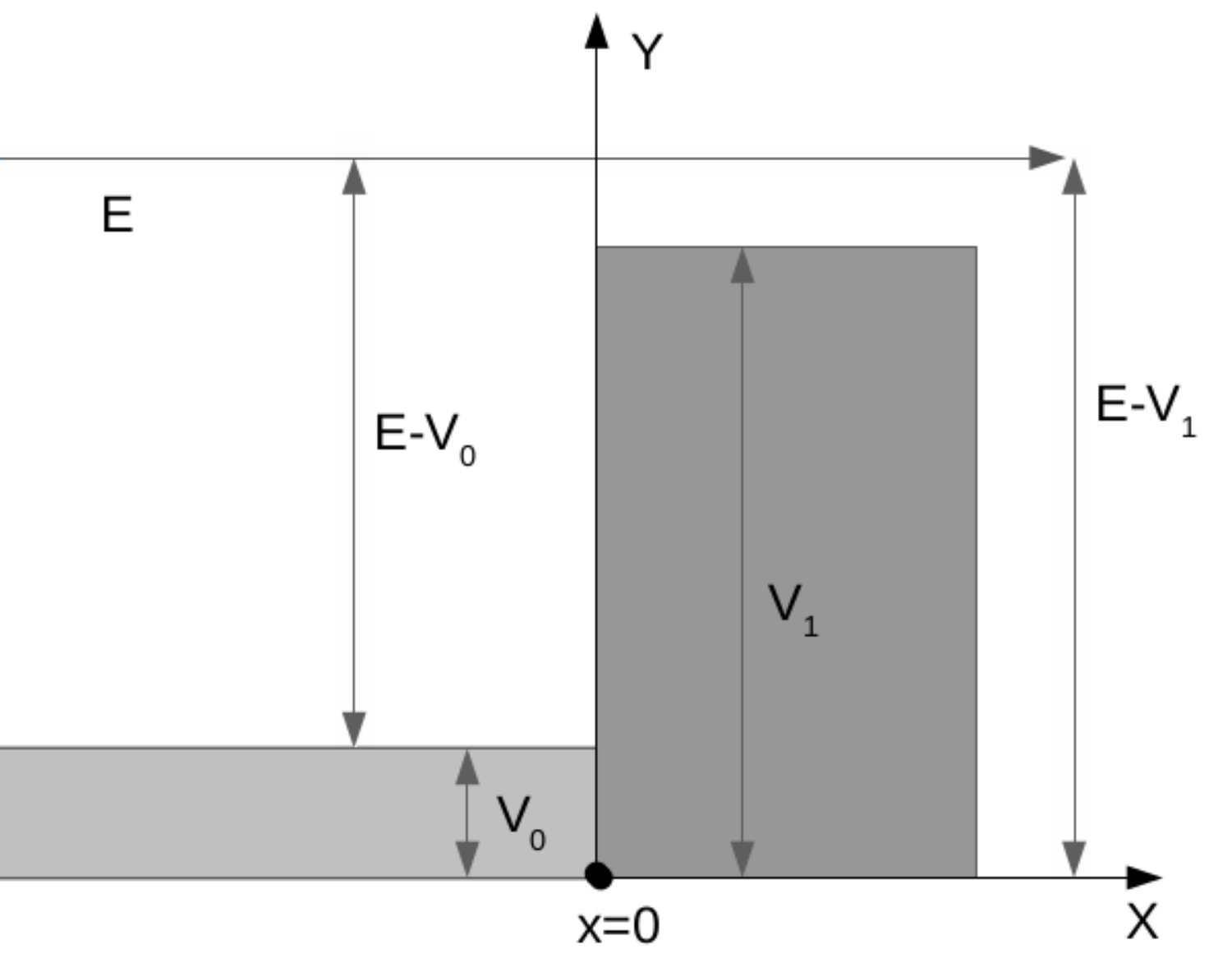}
  \caption{}
  \label{fig1b}
\end{subfigure}
\caption{(a) Particle (or ray of light) of energy $E$ in the region with potential $V_0$ is incident on a different potential barrier $V_1$ obliquely at an angle $\theta_0$. We assume $E>V_0, V_1$.}
\label{fig1}
\end{figure}

Due to the presence of the potential $V_0$ at $x<0$, the momentum of the particle (or a ray of light) represented by the wavefunction $\psi$ is: $k_0=\sqrt{\frac{2m(E-V_0 )}{\hbar^2}}$. At $x>0$, the momentum of the transmitted particle (or the light ray) is: $k_1=\sqrt{\frac{2m(E- V_1)}{\hbar^2}}$. The reflected component of the original wavefunction continues to be in the region of potential $V_0$, and has a momentum eigenvalue $k_0$ (with the sign of the $x$ component is reversed).

\subsection{Understanding of reflection and refraction}
Since $|{k_0}_x|$ and ${k_0}_y$ for the incident and reflected wavefunctions are the same, the angle of incidence must be equal to the angle of reflection. The translation invariance of the problem in the $Y$ direction demands:
\begin{equation}\label{Eq12}
 {k_0}_y={k_1}_y\\
\implies k_0\sin\theta_0=k_1\sin\theta_1
\end{equation}
Using the above expressions of momenta, we find:
\begin{equation}\label{Eq13}
 \sqrt{E-V_0}\sin\theta_0=\sqrt{E-V_1}\sin\theta_1
\end{equation}
At this point, we exploit the identifications made in the context of `F=ma Optics' and Eq.\eqref{Eq11}, that the total `energy' of the light rays is $0$, and the `potential energy' of light rays in a medium with refractive index $n$ is $-\frac{n^2}{2}$. Replacing these in Eq.\eqref{Eq13}, we find that it reduces to 
\begin{equation}\label{Eq14}
 n_0\sin\theta_0=n_1\sin\theta_1
\end{equation}
which is Snell's law, established from the scalar wave description of light in a purely algebraic manner. Students who are unfamiliar with Hamiltonian optics, but have a basic idea of Schr$\rm{\ddot o}$dinger's equation, can fathom this derivation.
\subsection{Estimation of coefficient of reflection}\label{REFL}
From figure~\ref{fig1}, the total wavefunction at $x<0$ is given as:

\begin{equation}\label{Eq15}
\psi_0(x,y)=e^{i{k_0}_xx+i{k_0}_yy}+\scriptr e^{-i{k_0}_xx+i{k_0}_yy}
\end{equation}
Where $\scriptr$ denotes the amplitude of reflection back into the region of space with potential $V_0$. On the other hand, the wavefunction at $x>0$ can be written as:
\begin{equation}\label{Eq16}
\psi_1(x,y)=\scriptt e^{i{k_1}_xx+i{k_1}_yy}=\scriptt e^{i{k_1}_xx+i{k_0}_yy}
\end{equation}
-where $\scriptt$ denotes the transmission amplitude and ${k_0}_y={k_1}_y$. The boundary conditions at $x=0$ are given by:\\ 
(a) $\psi_0(x=0)=\psi_1(x=0)$ and \\
(b) $\left(\frac{\partial\psi_0}{\partial x}\right)_{x=0}=\left(\frac{\partial\psi_1}{\partial x}\right)_{x=0}$.\\
The first condition yields:
\begin{align}\label{Eq17} 
    e^{i{k_0}_yy}+\scriptr\cdot e^{i{k_0}_yy} &= \scriptt e^{i{k_0}_yy}\nonumber\\
    \implies1+\scriptr &= \scriptt
\end{align}
The second condition implies:
\begin{align}\label{Eq18}
    i{k_0}_xe^{i{k_0}_yy}-\scriptr\cdot i{k_0}_xe^{i{k_0}_yy} &= \scriptt i{k_1}_xe^{i{k_0}_yy}\nonumber\\
    \implies{{k_0}_x - \scriptr\cdot{k_0}_x = \scriptt{k_1}_x}
\end{align}
From Eq.\eqref{Eq17} and Eq.\eqref{Eq18}, we can show that the reflection coefficient ($R=||\scriptr||^2$) can be expressed as:
\begin{align}\label{Eq19}
 R = \left|\left|\frac{{k_0}_x-{k_1}_x}{{k_0}_x+{k_1}_x}\right|\right|^2
\end{align}
In Eq.\eqref{Eq19}, ${k_0}_x=k_0\cos\theta_0=\sqrt{E-V}\cos \theta_0$ etc. Hence,
\begin{align}\label{Eq20}
 R &= \left|\left|\frac{k_0\cos\theta_0 - k_1\cos\theta_1}{k_0\cos\theta_0 + k_1\cos\theta_1}\right|\right|^2
 =\left|\left|\frac{\sqrt{E-V_0}\cos\theta_0 - \sqrt{E-V_1}\cos\theta_1}{\sqrt{E-V_0}\cos\theta_0 + \sqrt{E-V_1}\cos\theta_1}\right|\right|^2
\end{align}
Let us now see the implication of this discussion in the context of geometrical optics. If we take $E=0$ and $V_j=-\frac{n_j^2}{2}$ in accordance with our understanding of refraction, the quantity $R$ can be written as:
\begin{equation}\label{Eq21}
 R =\left|\left|\frac{n_0\cos\theta_0 - n_1\cos\theta_1} {n_0\cos\theta_0 + n_1\cos\theta_1}\right|\right|^2
 =\left|\left|\frac{n_0\cos\theta_0 - \sqrt{n_1^2-n_0^2\sin^2\theta_0}}{n_0\cos\theta_0 + \sqrt{n_1^2-n_0^2\sin^2\theta_0}}\right|\right|^2
\end{equation}
As expected, $R\rightarrow1$, when the incident angle $\theta_0\rightarrow90^o$, or if the condition for the total internal reflection $n_0\sin\theta_0= n_1$ is satisfied. We notice that Eq.\eqref{Eq21} is the same as Fresnel's equation of the coefficient of reflection for $s$-polarized light. The polarization of light in terms of electromagnetic theory is depicted in the following figure~\ref{fig2}.

\begin{figure}[ht]
\centering
\begin{subfigure}{.45\textwidth}
  \centering
  \captionsetup{justification=centering}
  \includegraphics[height=5.5 cm, width= 8.0 cm]{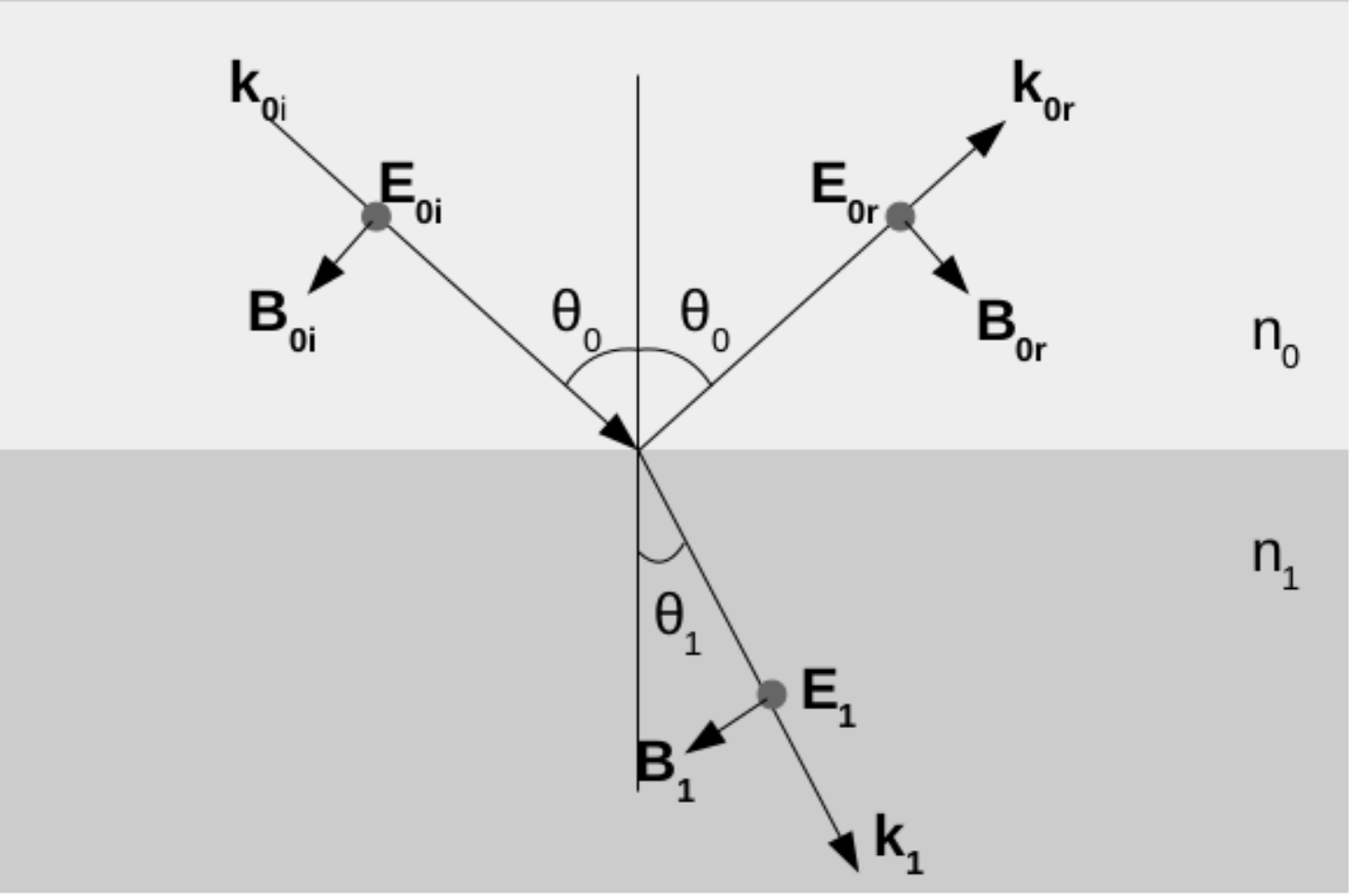}
  \caption{}
  \label{fig2a}
\end{subfigure}
\hspace{0.5 cm}
\begin{subfigure}{.45\textwidth}
  \centering
  \captionsetup{justification=centering}
  \includegraphics[height=5.5 cm, width= 8.0 cm]{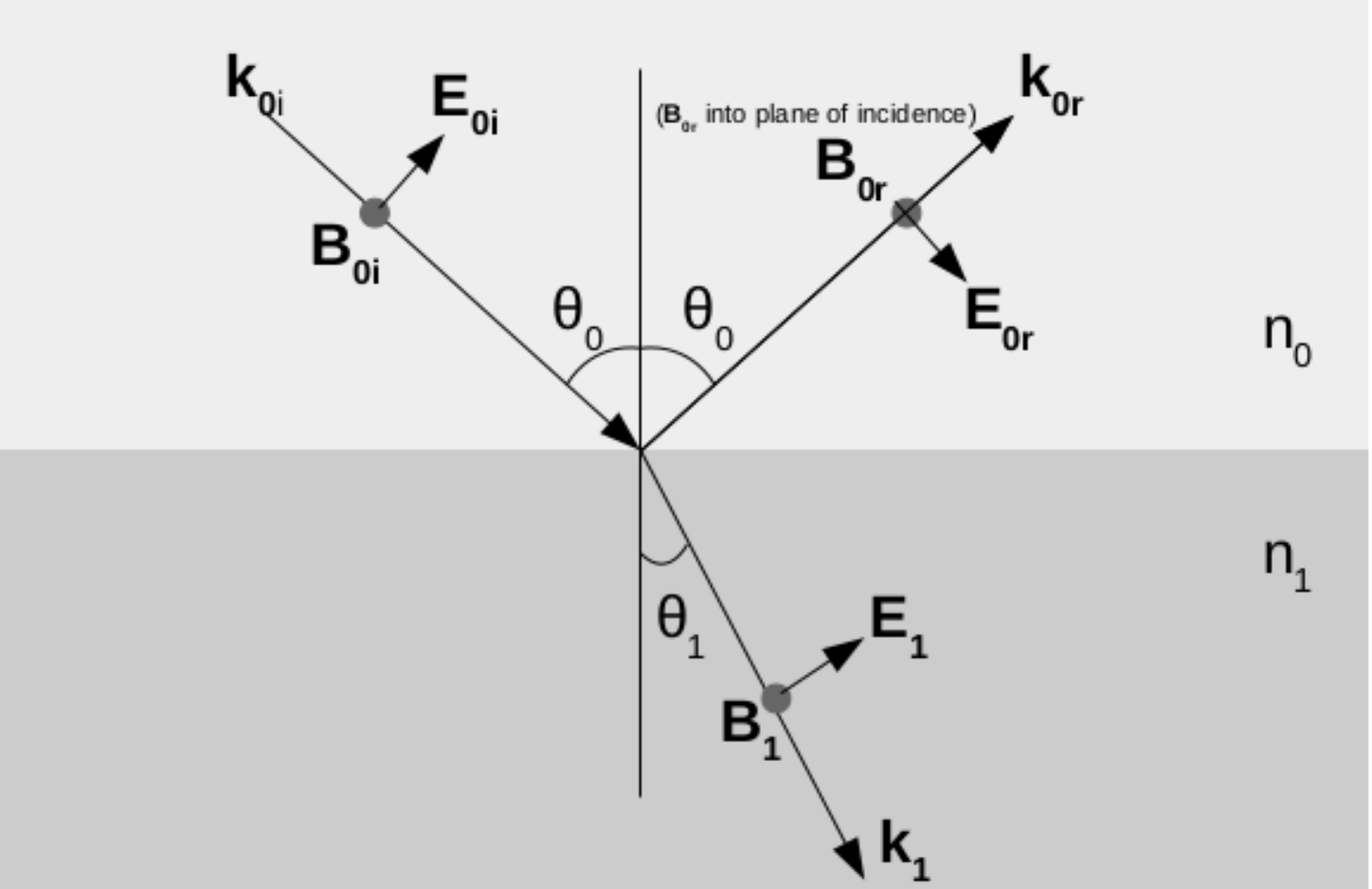}
  \caption{}
  \label{fig2b}
\end{subfigure}
\caption{Description of polarization of light in the context of incidence of light rays at the interface between two media with refractive indices $n_0$ and $n_1$; (a) $s-$polarized light: Electric field perpendicular to the plane of incidence, and (b) $p-$polarized light: Electric field in the plane of incidence.}
\label{fig2}
\end{figure}

This is expected, of course; but a curious student may wonder if there is a way to find the corresponding expression for $p$-polarized light. In fact, this formulation does not seem to lead to that expression. This must be the result of an incomplete description of the polarization of the light that is inherent in the ray picture. Gloge et al.~\cite{gloge1969formal} write ``we cannot expect that the total content of Maxwell's equations can be restored by our quantization concept, since the ray picture does not contain any information about the photon spin''. 

\section{Implication for physics pedagogy}
Commonly, the physics curriculum at the undergraduate level does not have significant interdisciplinary elements. However, this approach often brings out useful pedagogical insights. In this paper, we showed this using an example of a very well-known concept of optics. Along this journey, we used preliminary concepts of analytical mechanics, quantum mechanics, physical optics, and the electromagnetic theory; and it was rewarding in identifying the connection between the classical and quantum description of the light rays; in finding the relation between the plane wave solutions of the reduced wave equation, and the phasors; and in deducing the reflection coefficient. Even the inability to derive the reflection coefficient for the $p$-polarized light was pedagogically insightful because the ray picture does not have the connotation of polarization of light. This exercise hints towards the notion that the different topics are connected within themselves in a unique way. It is hoped that this article will help in the instruction of physics.

\bibliographystyle{unsrt}
\bibliography{ScalarLightWave}
\end{document}